\documentclass[useAMS,usenatbib]{mn2e}
\usepackage{graphicx,amsmath,multirow,amssymb}
\usepackage{natbib}
\newcommand{\comment}[1]{}

\def\simgt{\lower.5ex\hbox{$\; \buildrel > \over \sim \;$}}
\def\simlt{\lower.5ex\hbox{$\; \buildrel < \over \sim \;$}}

\title[Mg--Al--Si nucleosynthesis in massive AGB]
{A deep insight into the Mg--Al--Si nucleosynthesis in massive AGB and SAGB stars}
\author[P. Ventura, R. Carini and F. D'Antona]{P. Ventura$^{1}$\thanks{E-mail:
paolo.ventura@oa-roma.inaf.it (AVR)}, R.Carini$^{1,2}$ and F. D'Antona$^{1}$\\
$^{1}$INAF-Osservatorio Astronomico di Roma, Via Frascati 33, Monte Porzio Catone 00040, Italia\\
$^{2}$Dipartimento di Fisica, Universit\`a di Roma ``La Sapienza'', Italy}
\begin{document}

\date{Accepted, Received; in original form }

\pagerange{\pageref{firstpage}--\pageref{lastpage}} \pubyear{2002}

\maketitle

\label{firstpage}

\begin{abstract}
The stars in globular clusters are known to differ in their surface chemistry: the spectroscopic
investigations in the last decades outlined the presence of star--to--star differences in the
abundances of the light elements, up to aluminium (and possibly silicon), suggesting that
some stars were contaminated by an advanced proton--capture nucleosynthesis.
The AGB stars are one of the most promising candidates in producing the pollution of the 
intra--cluster medium, via the ejection of gas processed by Hot Bottom Burning, from which
new stellar generations are formed. This work is focused on the degree of nucleosynthesis
involving magnesium, aluminium and silicon that these sources may experience. 
The key ingredient to determine the degree of magnesium depletion, and the amount of aluminium that 
can be produced, is the rate of proton capture on $^{25}$Mg, forming $^{26}$Al; an increase in this 
cross-section by a factor 2 with respect to the highest value allowed by the NACRE compilation 
allows to reproduce the extent of the Mg--depletion observed, and is in qualitative 
agreement with the positive Al--Si correlation observed in a few clusters.
The main uncertainties associated with the macro-- and micro--physics input, are discussed and 
commented, and the comparison with recent spectroscopic results for the globular cluster showing 
some degree of Mg--Al anticorrelation and Al--Si correlation is presented. 
\end{abstract}


\section{Introduction}
The recent spectroscopic analysis of stars in Globular Clusters (GC) \citep{carretta09a, carretta09b}
have confirmed the results of the early investigations \citep{kraft}, i.e. the existence of 
star--to--star chemical differences, that define general abundance patterns, such as the well 
known O--Na and Mg--Al anticorrelations \citep{carretta06}. Such general trends, coupled with the 
absence of any spread in the heavy elements \citep{carretta09a}, and the constant C+N+O \citep{ivans99},  
indicate that the stars with the most anomalous chemistry are a second generation (SG), formed from gas 
contaminated by p--capture nucleosynthesis.
Among the proposed possible models, one of the most appealing hypothesis is that the intra--cluster medium 
(ICM) was polluted by the ejecta of intermediate mass stars during their Asymptotic Giant Branch (AGB) 
phase \citep{cottrell81}.
The most massive of these objects evolve rapidly compared to the typical age of
GCs, and expell gas that was mainly contaminated by Hot Bottom Burning (HBB), with
an advanced nucleosynthesis achieved at the bottom of their convective mantle, and the
consequent ejection of matter processed by hydrogen burning \citep{ventura01}.
The last few years have seen a great improvement in the theoretical description of the evolution 
properties of Super Asymptotic Giant Branch stars (hereinafter SAGB), i.e. objects whose
mass is in the range $6 \simlt M/M_{\odot} \simlt 8$\footnote{The range of mass involved depends on the
assumptions concerning the overshoot from the border of the convective core during the phase
of core hydrogen burning. In this work we assume a moderate extra--mixing; if the overshoot had
been neglected, the masses of SAGBs would be $8 \simlt M/M_{\odot} \simlt 10$}, 
that achieve an off-centre carbon ignition in
conditions of partial degeneracy, to end up their evolutionary history as White Dwarfs made up
of Oxygen and Neon \citep[see][and references therein]{siess06}. The recent compilation by
\citet{siess10}, in which a complete set of yields from SAGBs of various metallicities are
presented, allowed the inclusion of this class of objects together with the AGBs as possible
pollutors of the ICM. The terminology ``pollution by massive AGBs'' can be adopted to indicate the
coupled contamination of the ICM by AGB and SAGB stars. The importance of SAGB stars within the context 
of this scenario is outlined in the paper by \citet{annibale}, where the authors discuss the possibility 
that these stars provided the gas from which the first--born stars of the second generation formed, 
a possibility expected to work out only in the most massive clusters.

The yields by AGB stars are uncertain, as they depend critically on the inputs adopted
to calculate the evolutionary sequences, not known from first principles. \citet{vd05a} showed
that a major role on the predicted physical and chemical properties of these stars is 
played by the treatment of convection. When a high--efficiency modelling of convection is 
adopted \citep[i.e. the Full Spectrum of Turbulence developped by][]{cm91} HBB conditions
are achieved rather easily for masses exceeding 4M$_{\odot}$; the consequent rapid increase in the
luminosity \citep{blocker91} favours a large mass loss, thus a smaller number of Thermal Pulses 
(TPs), and a shorter duration of this evolutionary phase. Under these conditions, the surface
chemistry will reflect the HBB equilibria, with little contamination from the Third Dredge--Up
(TDU). The different treatment of the convective phenomenology is the main reason for the different
results obtained by the various research groups involved in this topic: AGB models in which
convection is modelled according to the Mixing Length Theory schematization achieve 
HBB conditions with more difficulty: \citet{fenner04} showed that their yields are not compatible 
with the chemical patterns observed in GC stars, because they show only a modest trace of HBB 
contamination. 

\cite{vd08, vd09} showed that when convection is modelled with the FST scheme and the mass loss
description accounts for the steep increase of $\dot M$ with luminosity as soon as HBB is ignited, 
the yields by massive AGBs are in agreement with the C--N anticorrelation observed in GC stars, and with 
the depletion of Oxygen detected. The O--Na anticorrelation can also be reproduced, provided that the upper 
limits for the proton capture reaction by $^{22}$Ne nuclei are chosen.

The modelling of SAGBs, for what concerns the HBB phenomenology, is less sensitive to the description 
of convection, because the core masses are so large that hot bottom burning conditions are easily reached; 
The yields by these stars are thus expected to show up the imprinting of HBB, although the extent of 
the nucleosynthesis experienced at the bottom of the convective mantle depends on the mass loss rate, 
that becomes the main reason of uncertainty \citep{vd11}; although the strength of the pulses is not
expected to be strong, the effects of the TDU can also be transported to the surface, provided that a
mass loss treatment with a soft dependence on the luminosity is adopted.
In comparison with the uncertain chemistry from AGBs, the yields by SAGBs 
can be reconciled more easily with the patterns observed in GC stars, the only constrain being the use of a 
mass loss rate steeply increasing with the luminosity \citep{vd11}. 

Provided the TDU is weak or made inefficient by the action of strong mass loss rate, the observed 
C-N and O-Na anticorrelations observed in GCs can be reproduced by the models; the situation is far
from being clear for what concerns the observed Mg-Al trend, with the possibile production of silicon.

The aim of the present work is to fill this gap, to understand whether the AGB and SAGB models, that show 
compatibility with the observed C--N, O--Na anticorrelations, can also reproduce the Mg--Al and Al--Si 
trends detected in some GCs. To achieve this task we investigate the modality of magnesium burning at the 
hot bottom of the convective envelope of massive AGBs. We analyze the thermal structure of these layers, 
to understand whether a significant depletion of the total magnesium, and a simultaneous enhancement in 
the surface aluminium, can occur, by comparing the burning timescale with the duration of the whole AGB phase. 
The possibility that some silicon is produced is also addressed. In agreement with previous investigations 
presented by our group, we also discuss the robustness of these predictions, and their sensitivity to the 
input--physics adopted.

\section{Mg-Al nucleosynthesis in massive AGBs}
\label{mgal}
A complete analysis of the nucleosynthesis of magnesium and aluminium is given in \citet{arnould}.
The investigation by \citet{kl03}, though limited to solar or slightly sub--solar metallicity, describes the 
different sites where the Mg--Al burning is active in AGB stars. The analysis is mainly focused on the
nucleosynthesis achieved within the 3$\alpha$ burning shell; the possible action of HBB is considered, but 
only their 6M$_{\odot}$ model of metallicity Z=0.004 (right--bottom panel of their Fig.~9) shows $^{24}$Mg 
consumption at the bottom of the envelope. The following paper by \citet{karakas06} is again focused on the 
nucleosynthesis in the helium--burning shell, and on the variation in the stellar yields due to the uncertainty 
in the cross--sections of the $\alpha$--capture reactions  by the magnesium isotopes. A physical context more 
similar to the present one can be found in \citet{dh03}, that studied the effects of HBB in massive AGBs of 
low metallicity. Their models outlined the possibility that when the temperature exceeds $\sim 10^8$K the 
surface $^{24}$Mg diminishes, with a consequent increase in the mass fractions of the heavier isotopes.
The Mg-Al nucleosynthesis in SAGB stars is studied in details by \citet{siess08}. Finally, the uncertainties 
of the relevant cross--section of the proton--capture reactions, and how they reflect on the yields from 
massive AGBs, were discussed in an exhaustive study by \citet{izzard07}.

The afore mentioned investigations showed that the synthesis of aluminium in AGBs is strictly 
associated to the activation of proton capture by $^{24}$Mg in the innermost regions of the 
outer convective zone, because this process starts a chain of proton capture reactions, that 
can ultimely lead to Al--production. The two isotopes $^{25}$Mg and $^{26}$Mg are also involved 
in this nucleosynthesis, that will change the overall content of magnesium.

As shown by \citet{dh03}, magnesium burning requires very large temperatures, exceeding $\sim 80$MK, 
because the cross-section of the $^{24}$Mg(p,$\gamma)^{25}$Al reaction, that gives origin to the whole chain, 
is extremely small at lower temperatures. The rate of $^{24}$Mg burning is the most steeply 
dependent on T, such that in a narrow range of temperatures this process switches from being practically 
inactive to become the fastest reaction of the whole chain. This is particularly interesting in the
case of massive AGB and SAGB stars, because they achieve very large
temperatures at the bottom of their convective zone, and are thus good candidates to 
produce gas which is depleted in the total magnesium content and enriched in aluminium.

\begin{figure}
\resizebox{1.\hsize}{!}{\includegraphics{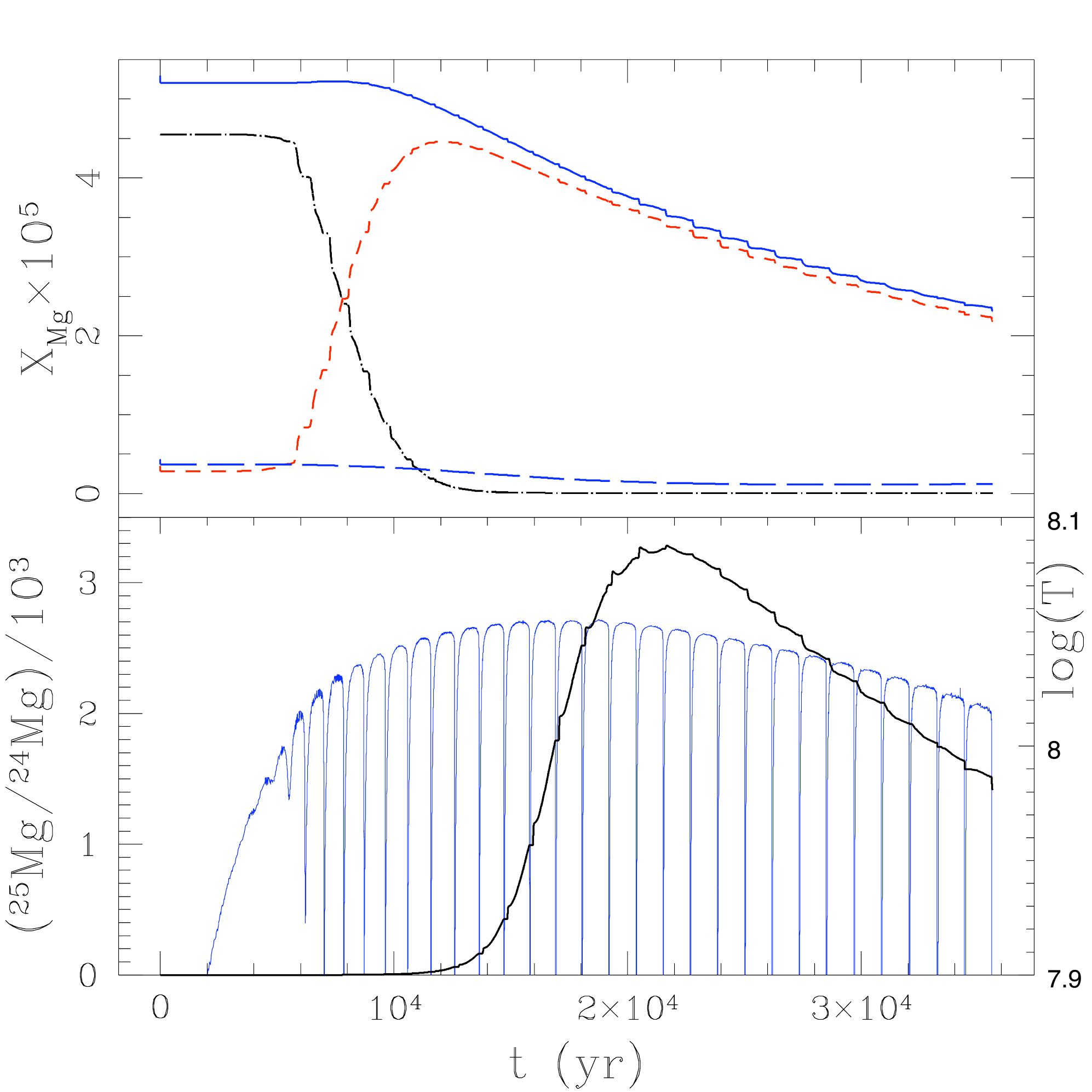}}
\caption{Top panel: variation with time (zeroed at the beginning of the AGB phase) of the surface 
abundances of the three magnesium isotopes and of the total magnesium in the standard
model of mass M=6M$_{\odot}$. Dot--dashed: $^{24}$Mg; short-dashed: $^{25}$Mg; Long--dashed: $^{26}$Mg; 
solid: $^{24}$Mg+$^{25}$Mg+$^{26}$Mg. Bottom panel: variation of the temperature reached at the bottom 
of the convective envelope (scale on the right) during the TPs, and ratio between the $^{25}$Mg and 
$^{24}$Mg abundances.}
\label{mag}
\end{figure}

The top panel of Fig.\ref{mag} shows the variation with time of the surface content of the 
three magnesium isotopes and of the total Mg in a model with initial mass 6M$_{\odot}$, 
evolved through the whole AGB phase. The behaviour of the temperature at the bottom of 
the convective mantle and of the $^{25}$Mg/$^{24}$Mg ratio is shown in the bottom panel.

Once the temperature exceeds $\sim 10^8$K, $^{24}$Mg burning begins, and proceeds with a time 
scale of the order of $\sim 500$ yr, much shorter than the duration of the AGB evolution, 
$\sim 3-4 \times 10^4$ yr. $^{24}$Mg has two production channels, i.e. the reactions
$^{23}$Na(p,$\gamma)^{24}$Mg and $^{27}$Al(p,$\alpha)^{24}$Mg, but this latter is
always negligible, because $^{27}$Al is destroyed preferentially via the alternative
channel that leads to the synthesis of silicon. The equilibrium value 
of $^{24}$Mg is therefore $\sim X(^{23}Na)\sigma (^{23}$Na+p) / $\sigma (^{24}$Mg+p), 
that, for the typical abundances of sodium within the envelope of AGBs, and in the range of 
temperatures of interest here, is of the order of $10^{-8}$. This is much smaller than the 
initial magnesium content ($X(^{24}Mg)_0 \sim 4\times 10^{-5}$), so magnesium is
destroyed until it reaches such a small equilibrium value after $\sim 10^4$yr (see dotted
line in Fig.~\ref{mag}). The exact equilibrium abundance is made uncertain by the scarce 
knowledge of the abundance of sodium, but this is not particularly relevant for the 
understanding of the behaviour of the overall magnesium and of aluminium.

The first product of $^{24}$Mg burning is $^{25}$Mg, which is seen to increase in the early
phases of AGB evolution (see the dashed line in the top panel of Fig.\ref{mag}), 
because the production
channel (proportional to X$(^{24}$Mg)$\sigma (^{24}$Mg+p)) largely exceeds the destruction rate
(proportional to  X$(^{25}$Mg)$\sigma (^{25}$Mg+p)). The surface content of $^{25}$Mg reaches a
maximum when $^{24}$Mg is sufficiently depleted, and is later destroyed, with a time
scale of the order $3\times 10^4$ yr. This time is comparable to the duration of the whole
AGB phase, thus $^{25}$Mg never reaches an equilibrium abundance in these thermodynamic
conditions; the $^{25}$Mg/$^{24}$Mg ratio (see bottom panel of Fig.\ref{mag}) increases with
continuity because $^{24}$Mg burning is by far more efficient, and then declines, when
the temperature at the bottom of the envelope begins to dimish due to the consumption of
the external envelope: this drop in T renders the destruction channel more competitive with
respect to the production rate, as a consequence of the different slope with T of the
two reactions (see Fig.~3 of \citet{siess08}).

The behaviour of $^{26}$Mg has a scarce impact on these results, because the decay of $^{26}$Al, 
which gives origin to $^{26}$Mg, is much slower ($\sim 10^6$ yr) than the 
$^{26}$Al(p,$\gamma)^{27}$Si reaction ($\tau \sim 600$ yr), thus favouring $^{26}$Mg destrucion, 
that occurs in a time scale of $\sim 5000$ yr. $^{26}$Mg is thus destroyed during
all the AGB phase (long-dashed line in Fig.\ref{mag}), and the overall magnesium
content will be given essentially by X($^{24}$Mg)+X($^{25}$Mg).

On the basis of these arguments, we conclude that $^{24}$Mg is burnt easily
at the bottom of the surface mantle of massive AGBs and SAGBs, and the overall depletion
of magnesium (hence, the amount of aluminium produced) will be determined by the 
velocity with which $^{25}$Mg, the first product of $^{24}$Mg burning, is destroyed.
Given the extremely small abundance of $^{26}$Mg, once $^{24}$Mg destruction occurs the total
magnesium will be given by $^{25}$Mg, thus the cross section of the reaction 
$^{25}$Mg(p,$\gamma)^{26}$Al is the key--quantity in determining the magnesium and 
aluminium content of the gas ejected by these stars.

\begin{figure}
\resizebox{1.\hsize}{!}{\includegraphics{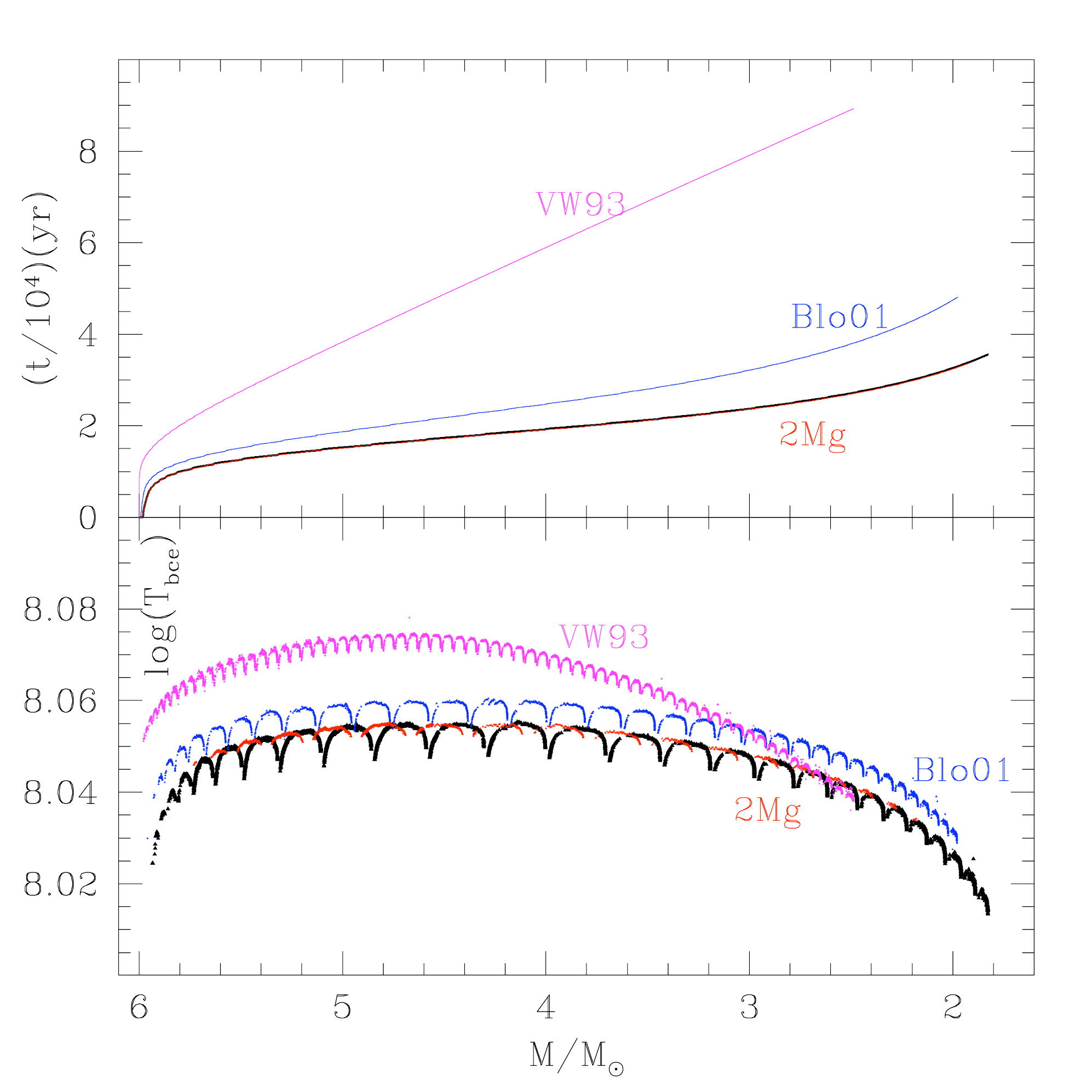}}
\caption{As a function of the total mass of the star we plot the temperature at the bottom of
the convective envelope when the H--burning shell is around its maximum efficiency during the TPs 
(bottom panel) and of the time (starting from the beginning of the AGB phase) of the 6M$_\odot$\ models 
(top panel). The acronyms close to the tracks are described in the text ---see Table 1.}
\label{fisica}
\end{figure}

\begin{figure}
\resizebox{1.\hsize}{!}{\includegraphics{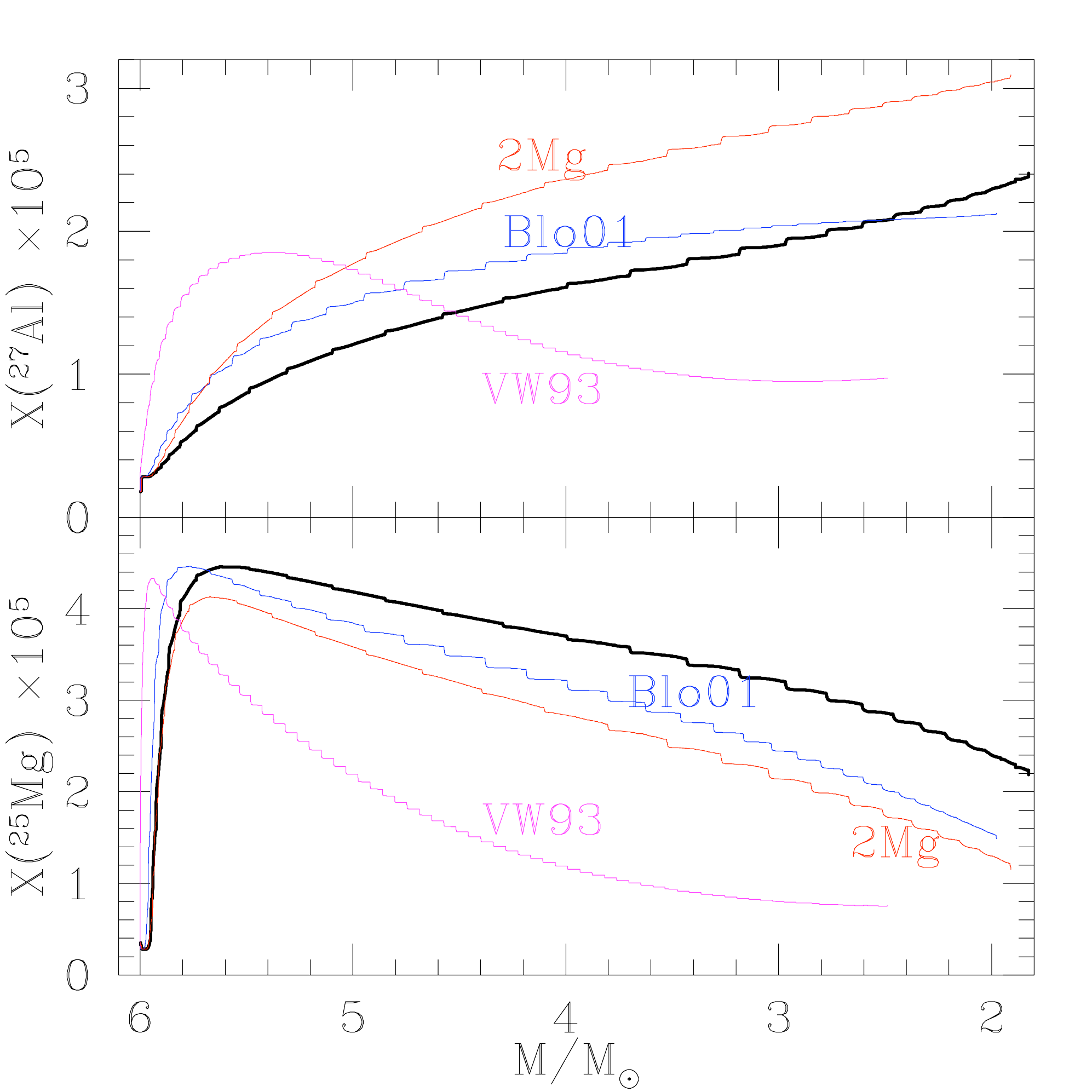}}
\caption{As a function of the total mass along the tracks, we show the surface abundances of $^{25}$Mg and $^{27}$Al during the AGB evolution of the four models discussed in the text.}
\label{mg25al27}
\end{figure}

The models producing the most extreme chemistries in our context are those with mass at the edge
between the AGB and the SAGB regime: we therefore focus on the 6M$_\odot$\ models, that are expected to 
provide the strongest contamination of the ICM \citep{vd11}. Our reference model, which we will refer to 
as ``standard'', was calculated with a chemistry typical of intermediate metallicity GCs, i.e. Z=0.001, 
and [$\alpha$/Fe]=+0.4; the mixture is taken form \citet{gs98}. 
Convection is addressed within the FST schematization, and mass loss is treated according to 
the \citet{blocker95} formulation, with the free parameter $\eta_R=0.02$, in agreement with the 
calibration given in \citet{ventura00}. To favour aluminium production, the upper limits for the
proton capture reaction by $^{25}$Mg nuclei were chosen; this is in agreement with \citet{ventura98},
where the interested reader may find all the details of the ATON evolution code, used in the present 
calculations.

Based on the results of the previous section, we focus our analysis of the role played by the choice 
of the cross--sections by comparing the results obtained when the rate of the reaction 
$^{25}$Mg(p,$\gamma)^{26}$Al is doubled with respect to the largest value allowed by the NACRE 
compilation; this is the model 2Mg. 

\begin{table*}
\caption{Relevant properties of AGB models}
\label{yields}
\begin{tabular}{ccccccccccc}
\hline
\hline
Model & $\tau_{AGB}$ & $\log(T_{\rm bce}^{\rm max})$ & $[^{16}$O/Fe] & 
[Na/Fe] & R(CNO) & [Mg/Fe] & $[^{27}$Al/Fe] & $[^{28}$Si/Fe] & 25/24 & 26/24 \\
\hline
standard  &  3.5$\times 10^4$  &  8.053 &  -0.40  &  0.31 & 0.97 &  0.27 & 1.04 & 0.44 & 27  & 1.4  \\
2Mg       &  3.5$\times 10^4$  &  8.053 &  -0.40  &  0.31 & 0.97 &  0.13 & 1.18 & 0.46 & 20  & 1.2  \\
Blo01     &  5$\times 10^4$    &  8.06  &  -0.53  &  0.15 & 0.94 &  0.18 & 1.00 & 0.49 & 38  & 1.6  \\
VW93      &  9$\times 10^4$    &  8.075 &  -0.63  &  0.05 & 1.65 & -0.11 & 0.83 & 0.62 & 117 & 4    \\
\hline
\end{tabular}
\end{table*}

The role of mass loss was investigated by calculating two models with $\eta_R=0.01$ (Blo01 model), 
and with a completely different treatment, i.e. the classic recipe by \citet{VW93} (VW93 model). 
The description of mass loss has a direct influence on the physical evolution of AGBs, because 
a higher mass loss favours a more rapid consumption of the whole envelope, with a shorter duration of 
the AGB phase \citep{vd05b}. Fig.\ref{fisica} shows the variation of the temperature 
at the bottom of the envelope (bottom panel) , and of the time counted from the beginning of the 
AGB phase (top), as a function of the total mass (decreasing due to mass loss) during the evolution. 
The choice of mass (instead of time) as abscissa allows a more direct understanding of the average 
chemistry of the gas ejected.

\begin{figure}
\resizebox{1.\hsize}{!}{\includegraphics{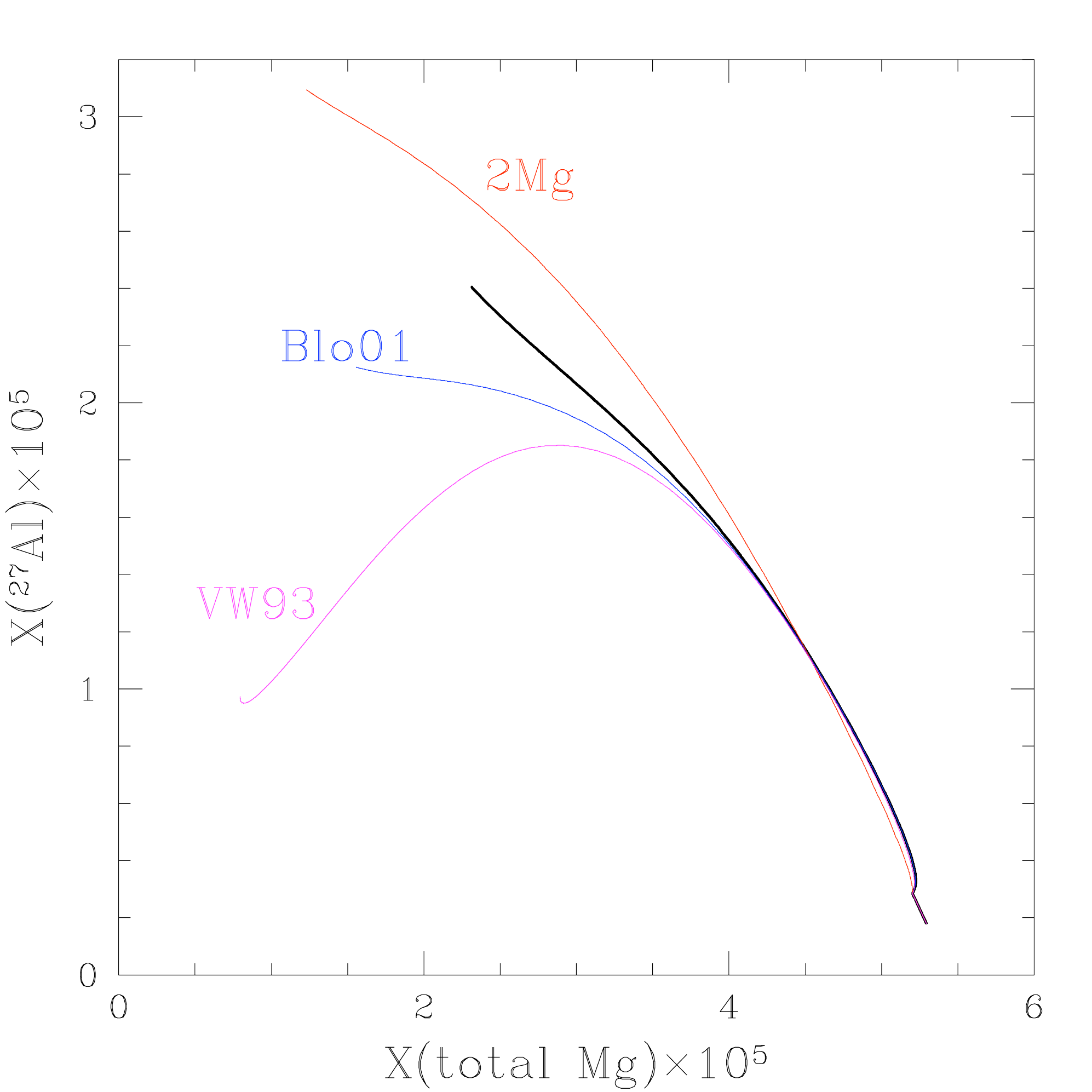}}
\caption{The simultaneous variation of the surface content of magnesium
and aluminium in the four models discussed in the text. The lines stop
when almost the whole envelope is lost.}
\label{mgal}
\end{figure}

The lines corresponding to the models standard and 2Mg cannot be distinguished. 
Models Blo01 and VW93 evolve more slowly, and, more important, attain higher temperatures 
at the bottom of their convective zone. 

The behaviour of the chemical species of interest is shown in Fig.\ref{mg25al27}.
For clarity reasons we only plot the behaviour of $^{25}$Mg and $^{27}$Al, as we have shown
that $^{24}$Mg is destroyed rapidly since the early TPs, and eventually the overall magnesium
content coincides with the $^{25}$Mg left in the envelope. The increase by a factor 2 in the 
$^{25}$Mg(p,$\gamma)^{26}$Al cross section favours a 25-30\% decrease in the overall magnesium.
The effect is not directly proportional to the increase in the cross-section used,
because $^{25}$Mg burning proceeds on time scales ($2-3\times 10^4$ yr) 
comparable to the total duration of this phase ($\sim 4\times 10^4$ yr, see Fig.\ref{fisica}).
The faster depletion of magnesium favours a larger production of aluminium, whose final
abundance is increased by $\sim 50\%$ compared to the standard case.

Mass loss modelling also influences magnesium: a slower
consumption of the envelope allows a stronger destruction of magnesium, that will survive in
smaller quantities within the external mantle. Fig.\ref{mg25al27} shows that the behaviour
of the surface abundance of magnesium is similar in all cases, and the steeper
decline of X($^{25}$Mg), particularly evident in the VW93 model, is a mere consequence of the 
smaller mass loss experienced. In the VW93 model the duration of the whole AGB phase becomes so long 
($\sim 10^5$ yr, as indicated by the VW93 line in Fig.\ref{fisica}) that $^{25}$Mg is destroyed 
until it reaches its final equilibrium abundance, as suggested by the flattening of the dotted--dashed 
line in Fig.\ref{mg25al27}. This suggests an upper limit to the total magnesium depletion that can be 
achieved within massive AGBs. 

The behaviour of aluminium is even more interesting. The $^{27}$Al produced in the envelope
increases both in the case when the $^{25}$Mg(p,$\gamma)^{26}$Al cross section is doubled or
with a smaller mass loss still in the Bl\"ocker framework (compare the 2Mg and Blo01
lines in Fig.\ref{mg25al27} with the standard counterpart), but decreases when the VW93 
recipe is adopted. The reason for this behaviour is the higher temperature reached at the 
bottom of the envelope (see also the bottom panel of Fig.\ref{fisica}), that favours 
the activation of $^{27}$Al burning with the consequent synthesis of some $^{28}$Si. This 
suggests that there is an upper limit to the amount of aluminium that can be produced within 
these stars, indipendently of the uncertainties associated to the cross--sections adopted.

The yields of the VW93 model can be compared with those of the same metallicity presented by
\citet{siess10}, based on the same description of the mass--loss mechanism. Such a comparison shows
that: a) we achieve a stronger Mg--depletion ($\delta$[Mg/Fe]=--0.5, to be compared to
$\delta$[Mg/Fe]=--0.15 by \citet{siess10}); b) we produce more Aluminium ($\delta$[Al/Fe]=+0.83, vs
$\delta$[Al/Fe]=+0.25). These differences can be understood on the basis of our choice to use the
upper limits for the $^{25}$Mg(p,$\gamma$)$^{26}$Al reaction, and to the higher magnesium in the
$\alpha$--enhanced mixture used in the present investigation, in comparison with the solar--scaled
mixture used by \citet{siess10}.

The results for the four 6~M$_\odot$\ evolutions are given in Table 1. The  duration of the AGB evolution  
($\tau_{AGB}$), the maximum temperature at the bottom of the convective envelope (T$_{\rm bce}^{\rm max}$), 
and the O, Na, Mg, Al and Si yields are given. The initial elemental abundances in the star are [X/Fe]=0.4 
for O and Mg and [X/Fe]=0 for Na and Si. 

Further information include the ratio of the total C+N+O abundances to the initial one (R(CNO)) and the 
average isotopic ratios $^{25}$Mg/$^{24}$Mg and $^{26}$Mg/$^{24}$Mg in the ejecta. These ratios exceed unity 
in all cases. The $^{26}$Mg/$^{24}$Mg ratio is less sensitive to the inputs used, because of the scarce 
impact of $^{26}$Mg in the overall magnesium content, and it deviates from unity only  in the VW93 case. 
In agreement with the previous discussions, we find that the $^{25}$Mg/$^{24}$Mg is much more sensitive to 
the modelling: this ratio decreases when the $^{25}$Mg(p,$\gamma)^{26}$Al cross--section is increased, whereas
it becomes higher when a smaller mass loss rate is used. The isotopic ratios are then 
$^{25}$Mg/$^{24}$Mg$\sim$20--40. Even larger values, as in other researchers' computations, are found in 
the VW93 model, where the ratio exceeds 100. This has been considered as a  major problem for all the models 
that attempt to explain the Mg--Al anticorrelation, as the analysis of stars in GCs NGC 6752, M71 and M13 
\citep{yong03, yong06} indicate that the ($^{25}$Mg+$^{26}$Mg)/$^{24}$Mg never deviates from unity, even in 
the most Al--rich objects. We do not discuss this problem any longer, but notice that it can find a simple 
solution in the framework of a dilution model \citep[e.g.][]{annibale}, as it is very easy to alter the 
high $^{25}$Mg abundance by dilution with pristine matter.

The variation of the surface content of aluminium in the four models discussed is shown as a function of the 
simultaneous change in the total magnesium in Fig.~\ref{mgal}. With respect to the standard model, the 2Mg 
model destroys more magnesium, and produces a larger amount of  aluminium: this is a mere consequence of the 
higher rate for the proton capture reaction by $^{25}$Mg nuclei, that moves the equilibrium towards aluminium. 
The models Blo01 and VW93 also destroy more magnesium, because the lower mass loss rate allows a more advanced 
nucleosynthesis, both thanks to the longer AGB evolution, and to the larger temperature reached at the bottom 
of the convective envelope. Unlike the 2Mg case, the total aluminium is decreased compared to the standard model, 
because the larger temperature favours the ignition of the $^{27}$Al(p,$\gamma)^{28}$Si, with a partial 
destruction of the aluminium synthesized. 
Thus we see that the amount of aluminium that can be produced by the AGB evolution is limited, on the lower 
boundary, by the HBB temperature and by the evolutionary time spent on the AGB (thus, on the mass loss rate) 
and, on the higher boundary, by the leakage into silicon, due to proton captures on aluminium.

Comparison of the magnesium yield of the standard track with the value published by \cite{vd11} alerted us to 
a mistake in the construction of those models, due to an incorrect value of the initial magnesium abundance. 
We then recomputed all the standard models and provide for clarity a full record of abundances in Table 2. 
Only the Mg and Al yields differ from the values of  \cite{vd11}.

\begin{table*}
\caption{Chemistry of the ejecta of SAGB models}
\label{yields}
\begin{tabular}{ccccccccccc}
\hline
\hline
Mass(M$_{\odot}$) & $\log(T_{\rm bce}^{\rm max})$ & $[^{16}$O/Fe] & 
[Na/Fe] & R(CNO) & [Mg/Fe] & $[^{27}$Al/Fe] & $[^{28}$Si/Fe] & 25/24 & 26/24 \\
\hline
  6.5M  &  8.064 &  -0.24  & 0.32 & 0.94 & 0.34  & 0.91 & 0.44 & 23.5 & 0.96 \\
  7.0M  &  8.079 &  -0.15  & 0.39 & 0.99 & 0.36  & 0.86 & 0.43 & 20.7 & 0.83 \\
  7.5M  &  8.10  &   0.01  & 0.67 & 1.12 & 0.365 & 0.65 & 0.42 & 8.7  & 0.39 \\
  8.0M  &  8.16  &   0.20  & 1.00 & 1.19 & 0.371 & 0.61 & 0.40 & 4.1  & 0.30 \\
\hline
\end{tabular}
\end{table*}

The revised magnesium and aluminium yield values of Table 2 for the standard tracks of 
6.5, 7, 7.5 and 8M$_\odot$\, are then plotted in Figures 6 and 7.

\section{Comparison with data}
\begin{figure}
\resizebox{1.\hsize}{!}{\includegraphics{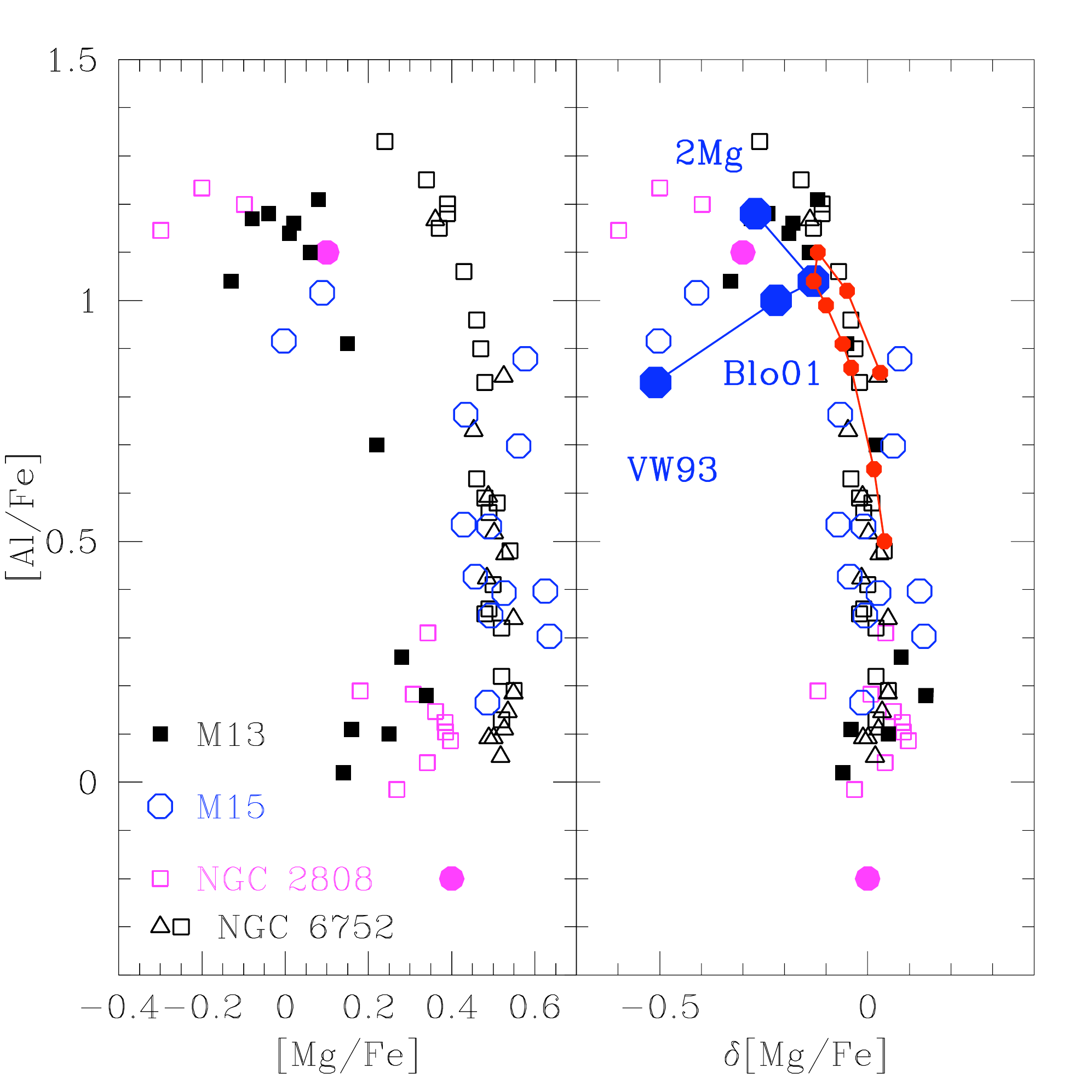}}
\caption{In the left panel, we show the [Mg/Fe], [Al/Fe] literature data. 
From Carretta et al. (2009a) we take the data for NGC6752 (open triangles), for M~15 (open circles) and for 
NGC~2808 (open squares). The extreme Mg abundances for three stars in NGC~2808 are from  Carretta et al. 
(2009b). From  Grundahl et al. (2002) we plot data for NGC~6752 (open squares), and from Sneden et al. (2004) 
the data for M~13 (full squares). The two turnoff stars measured by Bragaglia et al. (2010) in NGC~2808 are 
plotted as full (magenta) circles. In the right panel we plot the [Al/Fe] as a function of the magnesium 
variation $\delta$[Mg/Fe] for the different sets of data, and compare it to the theoretical variations 
expected from AGB and SAGB models. The $\delta$[Mg/Fe] values were obtained by applying to
stars belonging to the same cluster a constant shift to [Mg/Fe], so that the stars with the highest magnesium
have $\delta$[Mg/Fe]=0. Full large (blue) dots: results from the present investigation. Full 
small (red) points: AGB and SAGB yields by Ventura \& D'Antona (2009) and Table 2. The yields sequence 
goes from the 8M$_\odot$\  yield (lowest  point) through 7.5, 7, 6.5, 6.3, 6M$_\odot$\ (the point of the 
standard evolution), 5.5, 5, 4.5 M$_\odot$. }
\label{mgaloss}
\end{figure}

It is now more than 20 years that in GCs star--to--star variations in the surface contents of aluminium 
and (at a smaller extent) magnesium have been detected. The classic paper by \citet{norris95} outlined 
that the CO--weak  stars of  $\omega$ Cen, NGC 6397, NGC 6752 and M4 were all Al--rich, with enhancements  
up to more than a factor $\sim 10$. The question of a possible Mg depletion has been more debated. By 
examining a large sample of stars in M3 and M13,  \citet{cohen} claimed a little, if any, Mg depletion
in even the most anomalous stars, those  showing depleted oxygen and enhanced sodium; this conclusion was 
at odds with the earlier analysis by \citet{sneden04}. The recent compilation of data from many GCs 
presented by \citet{carretta09a} confirmed in most cases the existence of stars greatly enhanced 
($\sim 1$ dex) in their Al content, whereas the Mg--depletion remains within a factor $\sim 2$; the 
only exception are three very oxygen poor giants in NGC 2808, that show up a higher Mg reduction (see 
their Fig.7). Some clusters also exhibit a small Si--enhancement.
\cite{yong2008nitrogen} find that the nitrogen abundance in the giants of NGC~6752 is positively 
correlated with silicon, aluminium and sodium, and anticorrelated with oxygen and sodium. 
Another interesting result is the recent comparison \citep{bragaglia10} of abundances between two stars 
belonging to the reddest and to the bluer main sequence of NGC 2808. These two sequences, characterized 
by a difference in the helium content \citep{piotto07}, are expected to represent the original stellar 
population in the cluster. The measured abundances are [Mg/Fe]=+0.4, [Al/Fe]=--0.2 for the star in the red 
main sequence, and [Mg/Fe]=+0.1, [Al/Fe]=+1.1 for the bluer object. They are consistent with the 
hypothesis that stars in the blue main sequence formed from gas processed into the most massive AGBs, 
and is thus helium rich \citep{dantona2005}. If the blue main sequence formed directly from the massive 
AGB or SAGB ejecta \citep{annibale} it should also be Mg--poor and Al--rich, as found.

In Fig. ~\ref{mgaloss} we select some of the clusters for which Mg and Al data are available. 
In the left panel we plot the Mg data as they are provided in the references for each set given in the 
Figure caption. The data for stars with low Al abundances (probably representing the first generation stars) 
show a large systematic differences, possibly due to intrinsic magnesium differences from cluster to cluster, 
or to differences in the abundance derivations. In order to compare with models, we then deal with the magnesium 
variations, shown in the right panel.
The standard abundance variations, that is the abundances scaled by 0.4~dex for AGB and SAGB masses 
from 4.5 to 8M$_\odot$\ are shown as (red) smaller dots \citep[][and Table 2]{vd09}. 
Note that the abundances in the AGB and SAGB ejecta do not vary monotonically with mass, and that 
the Al (Mg) abundance reaches a maximum (minimum) extent around 6M$_{\odot}$, i.e. for the stars at 
the edge between the AGB and the SAGB regime. This effect, discussed in detail in \citet{vd11}, is 
due to the high mass loss experienced by the more massive objects, that favours a fast consumption of 
the whole envelope, before an extreme degree of nucleosynthesis is reached in the external mantle. 
The evolutions computed in this work are shown as big (blue) dots. We graphically see that the 
magnesium processing at 6M$_\odot$\ becomes larger, either by increasing the 
$^{25}$Mg(p,$\gamma)^{26}$Al cross section (2Mg dot) or by decreasing the mass loss rate (Blo01 
and VW93). 
The big dots indicate the range of values, expected for the most extreme magnesium depletion and
aluminum synthesis, that can be produced within AGBs and SAGBs, for various choices of the mass--loss
treatment and cross--sections.

The standard model provides an Al--enhancement in agreement with the Al abundances of the most Al rich stars, 
while its Mg--depletion, $\delta$[Mg/Fe]=--0.13, appears too small when compared 
to the most extreme data of M~15, NGC ~2808 and M~13. We see that both the increase in the 
$^{25}$Mg(p,$\gamma)^{26}$Al cross--section and the decrease in the mass loss rate allows a larger depletion 
of magnesium, in better agreement with 
the observations. The depletion of magnesium in the ejecta of the VW93 model is particularly 
strong ($\delta$[Mg/Fe]=--0.5), so it would allow to reproduce the smallest observed abundances. 
However, the yields predicted in this case could hardly reproduce the overall observed picture, 
because the Al--enhancement would be limited to 0.8 dex, and, more important, no sodium enhancement 
would occur (see Table 1). Few data would not agree with model 2Mg, and their abundance 
determinations must be carefully scrutinized before we draw conclusions. Notably, these stars include 
the three NGC~2808 stars studied by \cite{carretta09b}, having Mg depleted by 0.5~dex or more, 
while the blue main sequence star studied by \cite{bragaglia10} in the same cluster has a Mg 
abundance in good agreement with the standard or the 2Mg model.

\section{Any silicon synthesis in GLobular Clusters?}
The investigation by \citet{carretta09a} indicates a general spread in the silicon
abundances detected in the GC stars examined, and a positive
correlation between Al and Si. The two lowest panels of their Fig.~10, showing the 
stars belonging to NGC 2808 and NGC 6752, suggest a positive Al--Si trend, with a slope of
about 0.07 (i.e. an increase of 1 dex in the aluminium content is accompanied to a
0.07 dex increase in the measured silicon mass fraction). 

\begin{figure}
\resizebox{1.\hsize}{!}{\includegraphics{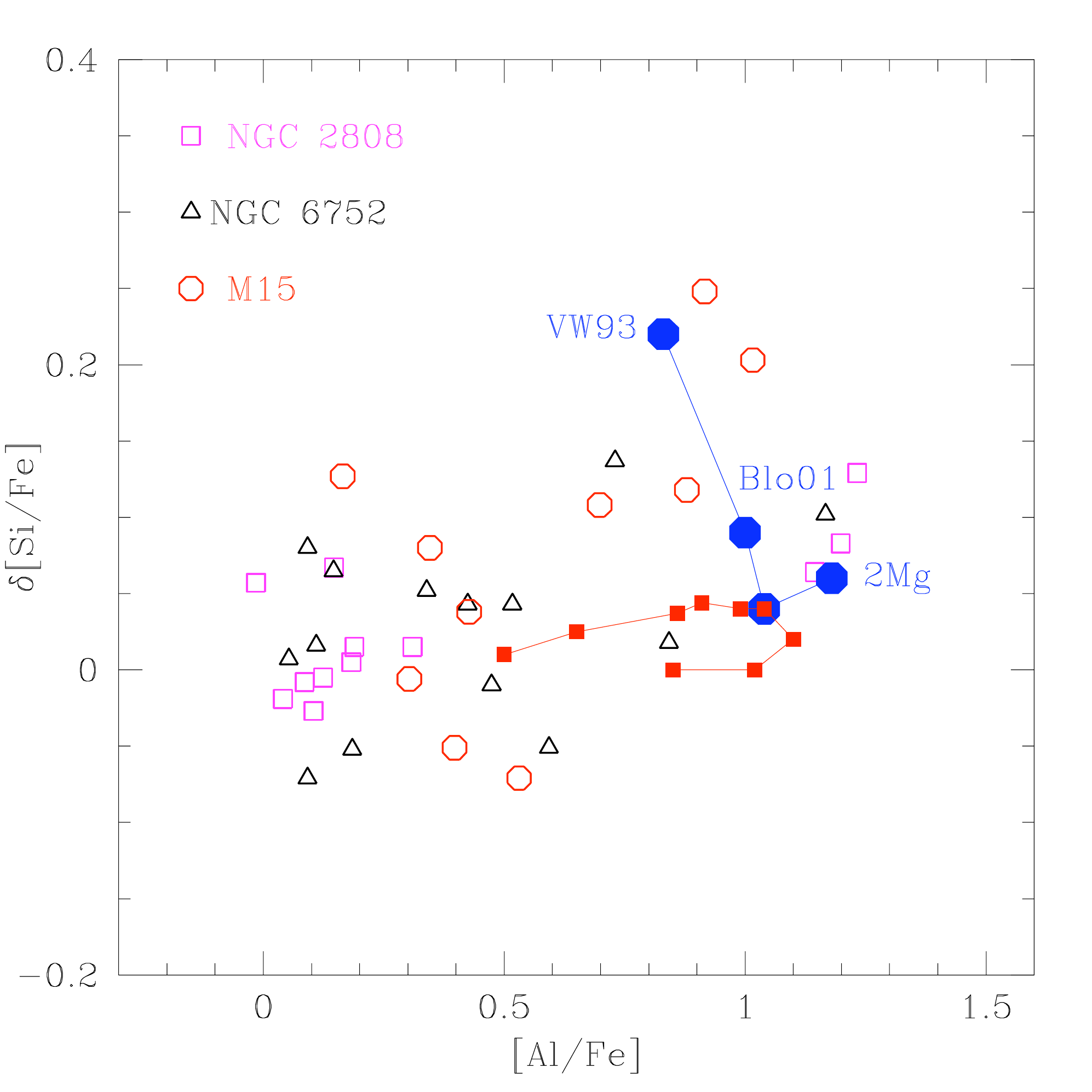}}
\caption{The same as Fig.~\ref{mgaloss}, but referring to the Al--Si plane. The silicon data and yields are normalized in order to show the [Si/Fe] variation. }
\label{alsioss}
\end{figure}

The SAGB and AGB silicon variations as a function of the aluminum abundance are plotted in Fig.~\ref{alsioss}, 
following the symbols of Fig.~\ref{mgaloss}. Data by \citet{carretta09b} for a few clusters are plotted. 
The stars with the largest Al--enhancement are also expected to be Si--enriched, because part of the aluminium 
produced is converted into Silicon. The area delimited by the large (blue) dots of the models computed in 
this work can be considered a measure of the degree of uncertainty of the strongest silicon production, 
coming form the scarce knowledge of the mass loss rate sufferd by these stars, and to the uncertainties 
associated to the relevant nuclear cross--sections. A larger production of Silicon is obtained when the 
rate of the p--capture process by $^{25}$Mg is increased (2Mg sequence) or when the mass loss rate is assumed 
to be smaller (Blo01 and VW93 sequences), the difference between the two cases being the opposite behaviour of, 
as discussed in the previous section. The increase in the rate of the  $^{25}$Mg(p,$\gamma)^{26}$Al reaction 
favours an increase in the silicon content  by $\delta$[Si/Fe]=0.06, which is possibly in better agreement 
with the data. The silicon synthesis is stronger in the Blo01 and VW93 models; in this latter case an increase 
$\delta$[Si/Fe]=0.22 is predicted, but these models are in contrast with the sodium abundances. We think 
that further observational work on the silicon abundances in the clusters showing the strongest chemical 
anomalies is needed before we can derive stronger constraints on the AGB and SAGB evolution.

\section{Conclusions}
We discuss the Mg--Al--Si nucleosynthesis at the bottom of the envelope of massive AGB
and SAGB stars, that experience strong HBB during their AGB evolution.
The results are compared with data for GC stars, that confirm the existence of a Mg--Al 
anticorrelation, and show the existence of a positive Al--Si trend. Our goal is to understand whether 
the same models that allow to reproduce the C--N and O--Na anticorrelations observed, can also account 
for these patterns involving Mg, Al, and Si. Similarly to the predictions regarding the O--Na trend, we 
find that when a description of mass loss with a steep dependence on the luminosity is used,
the most extreme chemistry, showing the maximum enhancement of aluminium and the strongest 
depletion of Magnesium, is reached within models of initial mass around 6M$_{\odot}$, 
that are at the edge between the AGB and the SAGB regime. SAGB stars, though
evolving at higher core masses, loose their mantle very rapidly, before a very
advanced nucleosynthesis can occur.

On the nuclear side, we confirm that $^{24}$Mg burning is started efficiently in all cases, 
and proceeds with time--scales so rapid that practically no $^{24}$Mg is left within the 
envelope. Given the small quantities of $^{26}$Mg produced, the key--quantity to determine 
the degree of magnesium depletion, and the amount of aluminium that can be produced, is the 
rate at which $^{25}$Mg nuclei capture protons, to form $^{26}$Al; an increase in this 
cross-section by a factor 2 with respect to the highest value allowed by the NACRE compilation,
although not supported by nuclear measurements, allows to reproduce better the extent of the 
Mg--depletion observed, and is also in better agreement with the slope of the positive correlation 
between Al and Si detected.

The extent of HBB experienced is also dependent on the treatment of mass loss. Use of a mass
loss rate scarcely dependent on the luminosity allows a more advanced nucleosynthesis at the
bottom of the convective envelope, that eventually leads to Al--destruction. Although these
models can deplete magnesium efficiently, they can hardly agree with the abundance patterns
observed, given that the Al-enhancement is limited to $\delta$[Al/Fe]=+0.8 and no sodium
production may occur. 

\section*{Acknowledgments}
The authors are indebted to the referee, L. Siess, for the careful reading of the
manuscript, and for the many comments and suggestions, that helped improved the
clarity and the quality of this work.
R.C. acknowledges financial support by the Observatory of Rome,
and MIUR/PRIN07 (Composizione chimica e popolazioni multiple negli ammassi globulari: 
osservazioni e modelli; CRA 1.06.07.05).

\end{document}